\documentclass{article}
\usepackage{amsfonts}

\newtheorem{claim}{Claim}

\newtheorem{lemma}{Lemma}

\newtheorem{theorem}{Theorem}
\newtheorem{corollary}{Corollary}

\newcommand{\remove}[1]{}
\newcommand{\qed}{\hfill\rule{2mm}{2mm}}

\newcommand{\bra}[1]{\langle#1\,|}
\newcommand{\ket}[1]{|\,#1\rangle}
\newcommand{\braket}[2]{\langle#1\,|\,#2\rangle}
\newcommand{\dyad}[2]{|#1\rangle\langle#2|}

\newcommand{\tr}{\mbox{\rm Tr}}

\newcommand{\vc}[1]{\mbox{\bf #1}}
\newcommand{\cons}{\sqsubseteq}

\newcommand{\bit}{\mathcal{B}}
\newcommand{\emptystring}{\xi}
\newcommand{\fail}{\mbox{\sf FAIL}}
\newcommand{\suc}{\mbox{\sf SUCC}}
\newcommand{\ssuc}{\mbox{\tiny\sf SUCC}}
\newcommand{\pv}{\preceq}
\newcommand{\dom}{\succeq}
\renewcommand{\ll}{\mbox{\sf LT}}
\newcommand{\rr}{\mbox{\sf RT}}
\newcommand{\conc}{;}
\newcommand{\sfidel}{\widetilde{F}}

\newcommand{\disc}{\mbox{\sf DIS}}

\newcommand{\todo}[1]{[{\bf TO DO:} \emph{{#1}}]}

\newenvironment{proof}{\begin{trivlist}
\item[\hspace{\labelsep}{\bf\noindent Proof: }]
}{\qed\end{trivlist}}
\newenvironment{sketch}{\begin{trivlist}
\item[\hspace{\labelsep}{\bf\noindent Proof's sketch: }]
}{\qed\end{trivlist}}

\title{Towards the Classical Communication Complexity of Entanglement 
  Distillation Protocols with Incomplete Information}
\author{
Andris Ambainis\thanks{Supported by NSF Grant CCR-9987845 and the
  State of New Jersey.}  {\sf $<$ambainis@ias.edu$>$}
\and
Ke Yang\thanks{
This research was sponsored by National Science Foundation (NSF)
grants CCR-0122581 and CCR-0085982. The views and conclusions
contained in this  document are those of the author and should not be
interpreted as  representing the official policies, either expressed 
or implied, of the NSF  or the US government.}
 {\sf $<$yangke@cs.cmu.edu$>$}}

\begin{document}
\maketitle

\begin{abstract} Quantum entanglement distillation protocols are
LOCC protocols between Alice and Bob that convert imperfect EPR
pairs, or, in general, partially entangled bipartite states into
perfect or near-perfect EPR pairs. The classical communication
complexity of these protocols is the minimal amount of classical
communication needed for the conversion. In this paper, we focus on
the communication complexity of protocols that operate with
\emph{incomplete information}, i.e., where the inputs are mixed
states and/or prepared adversarially.

We study 3 models of imperfect EPR pairs. In the measure-$r$ model,
$r$ out of $n$ EPR pairs are measured by an adversary;  in the
depolarization model, Bob's share of qubits underwent a
depolarization channel; in the fidelity model, the only information
Alice and Bob possess is the fidelity of the shared state.

For the measure-$r$ model and the depolarization model, we prove
tight and almost-tight bounds on the outcome of LOCC protocols that
don't use communication. For the fidelity model, we prove a lower
bound on the communication complexity that matches the upper bound
given by Ambainis, Smith, and Yang~\cite{ASY02}.

\remove{
over ``imperfect'' EPR pairs shared between Alice and Bob as input,
output perfect or near-perfect EPR pairs. 
In this paper, we study the communication complexity of these
protocols, that is, the minimal number of (classical) bits needed 
for Alice and Bob. 
We consider 3 models for imperfect EPR pairs. In the measure-$r$
model, $r$ out of $n$ EPR pairs are measured by an adversary (and thus
become ``corrupted''); in the
depolarization model, the qubits of Bob underwent a depolarization
channel; in the fidelity model, the only information Alice and Bob
possess is the fidelity of the shared state. 
For the first 2 models,
we prove tight and almost-tight bounds on the base case, i.e., how
well Alice and Bob 
can do if they are not allowed to communicate. 
For the fidelity model,
we prove a very  tight lower bound (up to an additive constant) on
the communication complexity. This lower bound implies the optimality
of the Random Hash protocol in~\cite{ASY02}. 
}
\end{abstract}

\section{Introduction}

\subsection{Entanglement Distillation
  Protocols} 

Quantum entanglement plays a central role in quantum information
theory. The phenomenon of having \emph{entangled} states separated by
space, is one of the quintessential features in  quantum mechanics. In
fact, one of the most fundamental problems in quantum information
theory is to understand entanglement. In particular, a very important
question is how to quantify entanglement: how do we measure the amount
the entanglement of a general bipartite state?

Not only is quantum entanglement conceptually interesting, it is very
useful ``in practice''. If Alice and Bob share EPR pairs~\cite{EPR35},
then they can perform teleportation~\cite{BBC+93}: Alice can transmit
an unknown  qubit to Bob by 
simply sending 2 classical bits. In this sense, shared EPR pairs
(paired with a classical communication channel) are
equivalent to a quantum channel. Furthermore, EPR pairs make
``superdense coding''~\cite{BW92} possible, where Alice can transmit
2 classical 
bits to Bob by only sending one qubit, provided that Alice and Bob
share an EPR pair a priori. However, qubits are prone to errors, and
EPR pairs may 
decohere and become imperfect. Can Alice and Bob
perform reliable teleportation and superdense coding if they share
imperfect EPR pairs?

Entanglement Distillation Protocols (EDPs) provide answers to both
questions mentioned above. Informally, EDPs are two-party
protocols that take imperfect EPR pairs (or general entangled states)
as input, and output bipartite states that are near-perfect EPR
pairs. During the execution of the protocol, both parties (denoted by
Alice and Bob) can perform local 
quantum operations (unitary transformations and measurements) on their
share of qubits, and communicate classical information. Alice
and Bob are not allowed to send qubits to each 
other. Protocols of this type are called ``LOCC
protocols'', standing for ``Local Operation Classical
Communication''. With EDPs, one can derive a quantity, namely the
``distillable entanglement'', for any bipartite state. The distillable
entanglement of a state  is the maximum number of EPR pairs Alice
and Bob can output using the optimal
EDP, which proved to be a very important quantity in measuring the
amount of entanglement for bipartite states. 
This answers the first question we mentioned
above. For the second question, Alice and Bob
can engage in an EDP to ``distill'' near perfect EPR pairs from
imperfect ones, and then use the distiled EPR pairs to perform
teleportation and superdense coding reliably.

\remove{

Despite the fact that EPR pairs are very important and useful, there
is a problem with them. Quantum states are prone to errors, qubits can
``decohere'' very easily, and make imperfect EPR pairs. To solve the
problem of imperfect EPR pairs, researchers have proposed Entanglement
Distillation Protocols (EDPs). 
Not only is entanglement distillation useful for producing perfect or
near-perfect EPR pairs, it provides considerable insights in
understanding quantum entanglement. One of the most important
questions in quantum information theory is ``how to measure
entanglement''.  
}

There have been numerous research efforts on entanglement distillation
protocols. We only list some of the most relevant work here.

To our knowledge, Bennett, Bernstein, Popescu, and Schumacher are the
first to consider the problem of producing EPR pairs from 
``less entangled'' states. In their seminal paper~\cite{BBP+96a}, they
gave a protocol that converts many identical copies of pure state
$\ket{\phi}=(\cos\theta\ket{01} +\sin\theta\ket{10})$ to perfect EPR
pairs. They call this process ``entanglement concentration''. In the
same year, Bennett, Brassard, Popescu, Schumacher, Smolin, and
Wootters~\cite{BBP+96b} studied the problem of ``extracting''
near-perfect EPR 
pairs from identical copies of mixed entangled states. This is the
first time that the notion ``entanglement purification protocols'' was
presented, which were renamed to ``entanglement
distillation protocols'' later~\footnote{The original name,
  ``entanglement purification protocols'', was considered less
  appropriate, since ``purification'' in quantum mechanics has other
  meanings.}. They 
also pointed out that EDPs can be used
to send quantum information through a noisy channel. Later,
Bennett, DiVincenzo, Smolin and Wootters~\cite{BDS+96} improved the
efficiency of the protocols in~\cite{BBP+96b} and proved a result that
closely related EDPs to quantum error
correcting codes, which is an alternative means to transmit quantum
information reliably through a noisy channel.

\remove{
Entanglement distillation protocols are not only useful, they also
provide insight to the understanding of quantum entanglement. 
Considerable research work has been done to understand the
``distillable entanglement''. Roughly speaking, the problem is as
follows: given a precise description of a bipartite quantum state, how
many EPR pairs can one ``extract'' by LOCC protocols? 
}

Horodecki, Horodecki, and Horodecki~\cite{HHH96, HHH99} and
Rains~\cite{R98a, R98b, R00} gave various asymptotic bounds on
distillable entanglement for arbitrary entangled states. They
considered the situation where $n$ identical copies of a state
are given as input to an LOCC protocol, which then outputs $m$ EPR
pairs. They 
studied the asympototic behavior of $m/n$ as $n$ approaches infinity.
Researchers also studied EDPs for a single copy of an arbitrary pure
state, see Vidal~\cite{V99}, Jonathan and Plenio~\cite{JP99},
Hardy~\cite{H99}, and Vidal, Jonathan, and Nielsen~\cite{VJN00}.
Much of the work was built on the result of majorization by 
Nielsen~\cite{N99a}, who is the first one that 
studied conditions under which one
pure state can be transformed into another one by LOCC.
All the work above assumes that Alice and Bob know the
explicit description of the state they share, and so they can act
\emph{optimally}.  

From another direction, researchers have studied EDPs
with \emph{incomplete information}, where
Alice and Bob don't know the exact state they share. 
The state is in a mixed state, or is prepared adversarially.
In this case we
cannot hope that Alice and Bob would act optimally. However, there 
still exist protocols that do reasonably well. Bennett
et. al~\cite{BBP+96b, BDS+96} studied the model where Bob's share
in the EPR pairs underwent a noisy channel, resulting in a
mixed state. They showed that their protocol would ``distill''
near-perfect  EPR pairs even when Alice and Bob don't have the
complete knowledge of the shared state. Under another circumstance,
``purity-testing protocols'' were studied
implicitly by Lo and Chau~\cite{LC99}, Shor and Preskill~\cite{SP00},
and later explicitly by Barnum, Cr{\'e}peau, Gottesman, Smith, and
Tapp~\cite{BCG+02}. Purity-testing protocols are LOCC protocols that
approximately distinguish the state of perfect EPR pairs 
from the rest states. Ambainis, Smith, and
Yang~\cite{ASY02} pointed out that purity-testing protocols are
indeed EDPs where Alice and Bob only know the \emph{fidelity} of
the state they share. Using constructions from~\cite{BCG+02},
Ambainis, Smith and Yang constructed a ``Random Hash'' 
protocol that produces $(n-s)$ EPR pairs of conditional fidelity at
least $1-2^{-s}/(1-\epsilon)$ on any $n$ qubit-pair input state of
fidelity  $1-\epsilon$. 
Their protocol would fail with probability $\epsilon$,
and the conditional fidelity of its output is the fidelity
\emph{conditioned on} that the protocol doesn't fail.

\subsection{Communication Complexity}

Classical communication complexity studies the minimal number of
(classical) bits 
needed to be transmitted between multiple parties in order to
collectively perform certain computation. Pioneered by Yao~\cite{Y79},
it is now a very rich field in theoretical computer science, and the
readers are refereed to~\cite{KN97} for more information. 

Quantum communication complexity, on the other hand, mostly studies
the minimal number of \emph{qubits} needed to be exchanged in order to
perform (quantum) computation. This field is also first studied by
Yao~\cite{Y93}, and now it is becoming one of the
central topics in quantum information theory. Numerous results have
emerged, and we refer the readers to~\cite{B01} for a nice survey, 
and~\cite{BBC+98,K01a, K01b,R02} for some important techniques and
results. 

However, another class of problem, namely the \emph{classical}
communication complexity for \emph{quantum} protocols, has being
largely ignored, until very recently. This class of problem is
concerned with the minimal number of classical bits needed to be
communicated to perform certain quantum computation.
An example is the classical communication complexity
for EDPs: one may ask ``how many bits do Alice and Bob need to
exchange in order to distill $n$ EPR pairs?''  One reason that
not many researchers pay too much attention to this problem might be
the conception that classical communication is ``cheap''
compared to quantum communication, and thus can be
ignored. However, as pointed by Lo and Popescu~\cite{LP99}, there are
situations where classical communication isn't ``that'' cheap that can
be justifiably ignored. One example is the super-dense
coding~\cite{BW92}: Alice and Bob can use $n$ qubits to transmit $2n$
bits of classical information, if they share $n$ EPR
pairs. Nevertheless, if it takes more than $n$ 
bits of classical communication to distill the $n$ EPR pairs, it would
totally destroy the purpose of super-dense coding. Furthermore, in the
study of LOCC protocols over quantum states, no quantum communication
takes place, and it is interesting to study the classical
communication complexity of these (quantum) protocols. The history of
classical communication complexity for quantum protocols can probably
traced back to the seminal paper by Bennett and Wienser~\cite{BW92},
which discussed teleportation and constructed a protocol that uses $2n$
classical bits to transmit $n$ qubits. However, this topic was largely
overlooked until by Lo and Popescu~\cite{LP99} and Lo~\cite{L99}. 
Lo and Popescu~\cite{LP99} discussed the
classical communication complexity of various protocols by
Bennett et. al.~\cite{BBP+96a}. They observed that the ``entanglement
concentration protocol'' in~\cite{BBP+96a} doesn't require any
classical communication, while the ``entanglement dilution protocol''
requires $O(n)$ bits of classical communication for producing $n$
copies of the ``diluted'' state. Lo and Popescu~\cite{LP99}
constructed a new dilution protocol that only uses $O(\sqrt{n})$ bits
of communication. This protocol was proven to be asymptotically
optimal by  
Hayden and Winter~\cite{HW02}, and Harrow and Lo~\cite{HL02}, who
proved matching lower bounds for general entanglement dilution
protocols. Lo~\cite{L99} studied the communication complexity for
Alice and Bob to jointly \emph{prepare} many copies of arbitrary
(known) pure states, and proved an non-trivial upper bound. All the
results above focus on a relatively simple situation, where the input
are $n$ copies of a known pure state, and only the asymptotic
results are known, i.e., the ratio of the amount of communication to
$n$ as $n$ approaches infinity.

\subsection{Our Contribution}

In this paper, we study the classical communication complexity of EDPs
with \emph{incomplete information}. In this setting, Alice and Bob
don't have the complete knowledge about the input state they share. 
Rather, the input state is a mixed state, or is adversarially
prepared. This is a natural extension to the simple model, where
Alice and Bob share a pure state. In fact, we argue that this is a
more ``realistic setting'': it is very hard, if not impossible, to
know \emph{precisely} which pure state a quantum system is in. Some
quite natural and commonly studied models of 
``noise'' in quantum state are probabilistic in
natural, and necessarily result in a mixed state. An
example is the depolarization channel. 
Furthermore, EDPs that work with adversarially prepared states have
the inherent worst-case behavior guarantee, and it more robust than
EDPs designed only specifically for some known pure states. It is,
therefore, very desirable to understand the communication complexity
of EDPs that work in this setting. 

We also study the \emph{precise} communication complexity of EDPs,
rather than their \emph{asymptotic} behavior. In fact, we try to
answer questions of the following fashion: ``On this particular 
input state class, how many bits of classical communication are needed
in order to just output a \emph{single} EPR pair with certain
quality?'' We feel that it is important to understand the
communication complexity in this case, where the requirement seems to
be \emph{minimal}. Interestingly, as we shall see later, the answer to
this minimal question already yields a lot of insights into the more
general problem, where 
Alice and Bob wish to generate EPR pairs of not only high quality, but
also of large quantity.

To the best of our knowledge, this is the first paper that studies
classical communication complexity of EDPs with incomplete
information, 
and also the first paper to address the precise communication cost,
rather than the asymptotic behavior. In fact, the only prior result
that studied classical communication complexity of
EDPs we are aware of is for the specific ``entanglement concentration
protocol''  by Bennett et. al.~\cite{BBP+96a}. As 
pointed by Lo and Popescu~\cite{LP99}, this protocol doesn't need any
classical communication. Notice that this particular protocol is the
first EDP that appears in literature, and works in perhaps the
simplest possible setting, where the 
input is a large copy of identical pure states. For all the related
work on classical communication complexity we are aware of, they all
work with a relative simple model.
In this model, Alice and
Bob try to convert many copies of some pure state $\ket{\phi}$ into
many copies of some other state $\ket{\psi}$. The fact that only many
copies of identical pure states are considered (and only asymptotic
results are needed) makes a lot of techniques
available, for example the Law of Large Numbers, the Central Limit
Theorem, and the conversion of  multiple-round protocols into
single-round protocols~\cite{BBP+96a,N99a,NC00,LP99,HL02,HW02}. These
techniques no longer work when we move to mixed input states
and ask for precise communication complexity.

As another motivation, we point out that, as EDPs are closely
related to Quantum Error Correcting Codes (QECCs), the
communication complexity of EDPs is closely related to the efficiency
of QECCs. Quantum error correcting codes are schemes to encode quantum
states redundantly, such that if part of the states are corrupted, one
can still recover the original encoded state. With QECC, Alice is able
to transmit quantum states reliably through a noisy quantum channel to
Bob. 
The readers are referred to~\cite{S95, S96, G97, NC00, P00} for more
discussions on QECCs. One of the central issues concerning QECCs is to
design QECCs that are efficient (i.e., has low redundancy) and robust
(i.e., tolerate a wide range of noise). As pointed by Bennett
et. al.~\cite{BBP+96b, BDS+96}, entanglement distillation
protocols can also be use to transmit quantum states reliably through
a noisy channel. Alice produces EPR pairs and sends Bob's share
through the noisy channel. Then Alice and Bob engage in an EDP to
``distill'' near-perfect EPR pairs. Finally Alice and Bob use the
shared near-perfect EPR pairs to perform teleportation and transmit
the quantum states reliably. From this point of view, entanglement
distillation protocols can be thought as ``interactive error
correcting protocols''. In fact, Bennett et. al.~\cite{BDS+96} proved
a relationship connecting QECCs and EDPs: they proved that QECCs and
1-way EDPs (where only Alice sends information to Bob and Bob doesn't
send anything back) are essentially
equivalent. From any 1-way EDP, one can derive a QECC with the same
parameter, and vise versa. They also showed that 2-way EDPs are more
powerful than QECCs in that there exists a noisy channel for which no
QECC is possible, but there exists 2-way EDPs that can transmit
information through this channel.  The communication complexity of
EDPs somewhat corresponds to the redundancy of QECCs. As in the case
of QECCs, it is therefore very desirable to construct EDPs of low
communication complexity that tolerate a high level of noise. In this
setting, the noise model is often adversarial or probabilistic, and
both precise and asymptotic results on communication complexity are
important.

We study EDPs in 3 different settings, corresponding to 3 different
models of ``imperfect'' EPR pairs. The first model is called the
\emph{measure-$r$} model. In this model, Alice and Bob originally
share $n$ perfect EPR pairs, and then $r$ out of these $n$ pairs are
measured in the computational basis. Each measured pair ends in a
mixed state ${1\over2}(\dyad{00}{00}+\dyad{11}{11})$, and becomes
disentangled. Alice and Bob have no information about which pairs are
measured and which are not, but they know $r$. In fact, we assume that
the $r$ measured pairs are \emph{adversarially} chosen.
This model is similar to the model
used in error correcting codes (both classical and quantum). 
The second model is called the \emph{depolarization} model. In this
model, $n$ perfect EPR pairs were produced by Alice, and then she
sends Bob's share of $n$ qubits to Bob through depolarization channel
of parameter $p$. In other words, each of Bob's qubits is 
left unchanged independently with probability $1-p$ and is replaced by
a completely mixed state with probability $p$. It is a typical model
for ``noisy channels'', and in particular was studied by 
Bennett et. al.~\cite{BBP+96b, BDS+96}. The third model is called the
\emph{fidelity} model. Here, Alice and Bob only know that the fidelity
of their shared state and perfect EPR pairs is $1-\epsilon$. Alice and
Bob don't have any other information about the state. This is the
model considered by Ambainis et. al.~\cite{ASY02}, where they called
it the ``general error'' model~\footnote{We feel that the name
  ``general error'' model isn't appropriate since this error model
  isn't the most general one. For example, it is not compatible with
  the previous error models.}.  

We obtain the following results: For the measure-$r$ 
model, we obtain a tight upper bound on the fidelity of
the output of protocols that don't use communication. More precisely,
we prove that in the measure-$r$ model, the maximal fidelity of
a protocol is at most $1-r/2n$, if no communication is
involved. Here we define the fidelity of a protocol to be the
worse-cast fidelity of the output of this protocol and the perfect EPR
pairs. This bound is tight in that we also present a (very simple)
protocol that achieves a fidelity of $1-r/2n$. Interestingly, the
proof seems quite non-trivial for this seemingly simple statement (and
the trivial protocol that matches the bound).
For the depolarization model, we obtain an almost-tight, similar
bound. We prove that in the depolarization model, the maximum
fidelity of a protocol is $1-p/2$, if no communication is
involved. This upper bound is almost tight, in that we also give
a (very simple) protocol that achieves $1-3p/4$. Both these 2 upper
bounds are for protocols that are only required to output 1
qubit-pair, which seems to be the minimal requirement for a ``useful''
EDP. For the fidelity model, we give almost tight (up to an additive 
constant) bounds on communication 
complexity of EDPs. More precisely, we prove that the maximal
conditional fidelity of an EDP of $t$ bits of communication is at most
$1-\epsilon\cdot p /2^{t+1}$, even if the EDP is only required to
output 1 qubit pair. Here $\epsilon$ is the fidelity of the input
state, and $p$ is the ``ideal success probability'', which is the
probability that the EDP succeeds with perfect EPR pairs (having
fidelity 1) as input. 
Therefore, to achieve a fidelity or $1-\delta$ on
the output, $\log(1/\delta)+\log(\epsilon\cdot p) -1$ bits of
classical communication is needed. Comparing the result
from~\cite{ASY02}, which constructed a protocol that uses
$\log(1/\delta)+\log(1-\epsilon)$ bits, our lower bound is tight up
to an additive constant. Here
we assume that both $\epsilon$ and $p$ are constant, which seems to be
the reasonable assumption. One interesting observation is that our
lower bound was 
proven for protocols what only output 1 qubit pair, while the matching
upper bound is from a protocol that outputs many qubits (in fact, in
the usual setting, the protocol outputs all but logarithmically number
of input qubit pairs). This seems to indicate that the communication
complexity is \emph{oblivious} of the \emph{yield} of the EDPs with
respect the fidelity model. This
fact is quite surprising, since it is definitely not the case for
QECCs.

All the proofs in out paper are from first principles and don't
involve very complex analysis. Some techniques used in this paper
would be interesting by themselves: in fact, as we pointed out
earlier, the old techniques don't work any more in our setting, when
mixed states and studied and we are interested in the precise
communication complexity. Therefore, we need to use new techniques,
among which are an alternative definition on fidelity, which proved
very useful in proving the first 2 bounds, and an observation on the
``splitting'' of mixed states during communication, which is useful to
prove the lower bound for the fidelity model.

\subsection{Outline of the Paper}

In Section~\ref{sec:notation}, we present some notations and
definitions to be used in the rest of  the paper. We prove a lower
bound for the measure-$r$ model in Section~\ref{sec:measure-r}. We
prove a lower bound for the fidelity model in
Section~\ref{sec:depolar}. In Section~\ref{sec:fidelity} we prove the
lower bound for the fidelity model. We conclude the paper in
Section~\ref{sec:conclusion}. Some proofs are postponed to the
Appendix.

\section{Notations and Definitions}
\label{sec:notation}
All logarithms are base-2. We identify an integer with the 0-1 vector
obtained from its binary 
representation. For a vector $v$, we write $v[j]$ to denote its $j$-th
entry. For 0-1 vector $x$, we denote its
\emph{Hamming weight} by $|x|$, which is the number of 1's in $x$. We
define $\bit=\{0,1\}$, and naturally $\bit^n = \{0,1\}^n$. For binary
strings $x$ and $y$, we use $x\conc y$ to denote the
\emph{concatenation} of these 2 strings.

Throughout the paper we are interested in finite, bipartite,
symmetric quantum systems shared between Alice and Bob. We identify a
``ket'' $\ket{\phi}$ with a unit column vector. We assume
there exists a canonical computational basis for any finite Hilbert
space of dimension $N$, and we denote it by
$\{\ket{0}, \ket{1}, ..., \ket{N-1}\}$. We use
superscripts to indicate which ``side'' a qubit or an operation
belongs to. For example, a general bipartite
state $\ket{\varphi}$ can written as
$\ket{\varphi} = \sum_{i,j}\alpha_{ij}\ket{i}^A\ket{j}^B$. 

There are 4 \emph{Bell states} for a pair of qubits shared between
Alice and Bob, and we denote them as follows:
\begin{eqnarray}
\Phi^+ & = & {1\over\sqrt{2}}(\ket{0}^A\ket{0}^B+\ket{1}^A\ket{1}^B) \\
\Phi^- & = & {1\over\sqrt{2}}(\ket{0}^A\ket{0}^B-\ket{1}^A\ket{1}^B) \\
\Psi^+ & = & {1\over\sqrt{2}}(\ket{0}^A\ket{1}^B+\ket{1}^A\ket{0}^B) \\
\Psi^- & = & {1\over\sqrt{2}}(\ket{0}^A\ket{1}^B-\ket{1}^A\ket{0}^B) 
\end{eqnarray}

We denote the state $(\Phi^+)^{\otimes n}$, which represents $n$
perfect EPR pairs, by $\Psi_n$. We also abuse the notation to use
$\Psi_n$ to denote \emph{both} the vector $\Psi_n$ and its density
matrix n$\dyad{\Psi_n}{\Psi_n}$, when there is no danger of
confusion. 

A quantum state is \emph{disentangled} if it is of the form
$\ket{\psi}^A\otimes\ket{\psi'}^B$.  Any other pure state is {\em
  entangled}. A mixed state $\rho$ is disentangled if
and only if it is equivalent to a state that is a mixture of 
disentangled pure states.
Any other mixed state is entangled.

The Pauli Matrices $X$, $Y$, and $Z$ are unitary operations over a
single qubit defined as
\begin{eqnarray*}
X(\alpha\ket{0} + \beta\ket{1}) & = & \beta\ket{0}+\alpha\ket{1} \\
Y(\alpha\ket{0} + \beta\ket{1}) & = & i\beta\ket{0} - i\alpha\ket{1}\\
Z(\alpha\ket{0} + \beta\ket{1}) & = & \alpha\ket{0} - \beta\ket{1}
\end{eqnarray*}
We use $I$ to denote the identity operator. 

For a unitary operator $U$, we can write it in a matrix form under the
computational basis. Then we define its \emph{conjugate}, $U^*$, to
the entry-wise conjugate of $U$. Clearly $U^*$ is still a unitary
operation. 

An \emph{error model} is simply a set of bipartite (mixed)
states, and is often denoted by $\mathcal{M}$. We say a state $\rho$
is \emph{consistent} with $\mathcal{M}$, if $\rho\in\mathcal{M}$.

\subsection{Fidelity}

For two (mixed) states $\rho$ and $\sigma$ in the same Hilbert space
their \emph{fidelity} is defined as

\begin{equation}
\label{eqn:def-fidelity-orig}
F(\rho, \sigma)=\tr^2(\sqrt{\rho^{1/2}\sigma\rho^{1/2}}).
\end{equation}
Notice we are using a different definition as in [NC00], where the
\emph{square root} of (\ref{eqn:def-fidelity-orig}) is used.

If $\sigma = \dyad{\varphi}{\varphi}$ is a pure state,
the definition simplifies to
\begin{equation}
\label{eqn:def-fidelity-pure}
F(\rho, \dyad{\varphi}{\varphi})=
\bra{\varphi}\rho\ket{\varphi}
\end{equation}

A special case for the fidelity is when $\ket{\varphi}=\Psi_n$ for
some $n$, such that $\rho$ and $\Psi_n$ have the same dimension.
In this case, we call the fidelity of $\rho$ and $\ket{\varphi}$ the
\emph{fidelity  of state $\rho$}, and the definition simplifies to:
\begin{equation}
\label{eqn:def-fidelity-new}
F(\rho) = \bra{\Psi_n}\rho\ket{\Psi_n}
\end{equation} 

We are often interested in the fidelity of 2 states of unequal
dimensions. In particular, we are interested in the fidelity of a
general bipartite state $\rho$, and the Bell state $\Phi^+$. This
coincides with the definition of fidelity when $\rho$ has dimension
2. When $\rho$ has a higher dimension, we define its \emph{base
  fidelity} to be the fidelity of the state obtained by tracing out
all but the first qubit pair of $\rho$. We denote the base fidelity of
$\rho$ by $\sfidel(\rho)$.

\remove{
A more special case is when one of the states is the Bell state
$\Phi^+$. We write this as
 $\sfidel(\rho) = F(\rho, \Phi^+)$, and call it the \emph{base
   fidelity} of state $\rho$. We further
extend this definition to states of higher dimension. 
For a bipartite state $\rho$ of dimension more than 2, $\sfidel(\rho)$
is defined to be the fidelity of the state obtained by tracing out all
but the first qubit pair.
}

It is easy to verify that the fidelity is linear with respect to
ensembles, so long as one of the inputs is a pure state, as in
the following claim.
\begin{claim}
\label{claim:linear-fidelity}
If $\rho$ is the density matrix for a mixed state that is an ensemble
$\{p_i,\;\ket{\phi_i}\}$, and  $\sigma$ is the density matrix of a
pure  state, then we have
$F(\rho, \sigma) = \sum_{i}p_i\cdot F(\dyad{\phi_i}{\phi_i}, \sigma)$.
\qed
\end{claim}

The fidelity is also monotone with respect to trace-preserving
operations~\cite{NC00}

\begin{claim}
\label{claim:fidelity-monotone}
For any states $\rho$ and $\sigma$ and any trace-preserving
operator $\mathcal{E}$, we have
$F(\mathcal{E}(\rho), \mathcal{E}(\sigma))\ge F(\rho, \sigma)$.
\qed
\end{claim}

One useful fact about fidelity is that any completely
disentangled state has base fidelity at most $1/2$.
\begin{lemma}
\label{lemma:disentangle-fidelity}
If $\rho$ is a completely disentangled state, then
$\sfidel(\rho)\le 1/2$. 
\end{lemma}
\begin{proof}
By the definition of base fidelity, we may assume that $\rho$ has
dimension 2.
By Claim~\ref{claim:linear-fidelity}, we only need to consider the
case that $\rho$ is a pure state $\dyad{\phi}{\phi}$. 
Since $\ket{\phi}$ is disentangled, we may write it as
$$\ket{\phi} = (\alpha_0\ket{0}+\alpha_1\ket{1})\otimes
(\beta_0\ket{0}+\beta_1\ket{1})$$
Then a direction calculation reveals that
\begin{eqnarray*}
\sfidel(\dyad{\phi}{\phi}) & = & 
{1\over2}\left|\alpha_0\beta_0 + \alpha_1\beta_1\right|^2 \\
& = & 
{1\over2}\left(|\alpha_0|^2|\beta_0|^2 + |\alpha_1|^2|\beta_1|^2 +
\alpha_0\beta_0\alpha_1^*\beta_1^* +
\alpha_0^*\beta_0^*\alpha_1\beta_1\right)\\
& \le & {1\over2}\left(|\alpha_0|^2|\beta_0|^2 +
  |\alpha_1|^2|\beta_1|^2 +
|\alpha_0\beta_1^*|^2 + |\alpha_1\beta_0^*|^2 \right)\\
& = & {1\over2}(|\alpha_0|^2+|\alpha_1|^2)(|\beta_0|^2+|\beta_1|^2)\\
& = & {1\over2}
\end{eqnarray*}
\end{proof}

\subsection{Entanglement Distillation Protocols}

We give a detailed description on  entanglement distillation protocols
discussed in our paper.
We often denote an entanglement distillation protocol by
$\mathcal{P}$. The protocol starts with a  mixed state $\rho$ shared
between Alice and Bob. Alice and Bob can have their private ancillary
qubits, originally initialized to $\ket{0}$. A protocol is either
deterministic or probabilistic. For \emph{deterministic}
protocols, Alice and Bob don't share any initial random bits; for
\emph{probabilistic} protocols, Alice and Bob share a (classical)
random string. We say a protocol $\mathcal{P}$ is a $t$-bit protocol,
if there are $t$ bits of (classical) communication during the
protocol.  We don't allow protocols to have any initial entanglement
as auxiliary inputs, and neither do we allow quantum channels between
Alice and Bob.

An the end of a protocol, both parties output $m$ qubits, which form
the output of the protocol. 
In addition, Alice also outputs a special symbol
(either a $\suc$ or a $\fail$). The \emph{success probability} of a
protocol ${\cal P}$ over an input state $\rho$ is the probability that
Alice outputs $\suc$ at the end of the protocol, and we write this as
$P^{\ssuc}_{\mathcal{P}}[\rho]$. The \emph{ideal 
  success probability} of a protocol ${\cal P}$ is its success
probability over input $\Psi_n$. We say a protocol is \emph{ideal}, if
its ideal success probability is 1. If $\sigma$ is the density matrix
of the output of protocol $\mathcal{P}$ on input $\rho$, we write it
as $\mathcal{P}(\rho) = \sigma$. If $\tau$ is the density matrix of
the output of protocol $\mathcal{P}$ on input $\rho$, \emph{conditioned
  on} that Alice outputs $\suc$, then we call $\tau$ the
\emph{conditional output} of protocol $\mathcal{P}$, and write this as
$\mathcal{P}^c(\rho)  = \tau$.

For an entanglement distillation protocol $\mathcal{P}$, we define its
\emph{fidelity} with respect to an error model ${\cal M}$, denoted by
$F_{\mathcal{M}}(\mathcal{P})$, to be the 
minimal fidelity of its output over all input state consistent with
$\mathcal{M}$. In other words,
\begin{equation}
F_{\mathcal{M}}(\mathcal{P}) =
\min_{\rho\in\mathcal{M}}F(\mathcal{P}(\rho)) 
\end{equation}
Similarly, we define the \emph{conditional fidelity} to be the
minimal fidelity of its conditional output, denoted by
$F^c_{\mathcal{M}}(\mathcal{P})$:
\begin{equation}
F^c_{\mathcal{M}}(\mathcal{P}) =
\min_{\rho\in\mathcal{M}}F(\mathcal{P}^c(\rho)) 
\end{equation}

When the error model ${\cal M}$ is clear from the context, it is
often omitted.

\remove{

one party sends a special symbol (either a
$\suc$ or a $\fail$) to the other party. If $\suc$ is sent, we say the
protocol \emph{succeeds}, and both party output $m$ qubits. 

A protocol is
either absolute or conditional. 
At the end of an \emph{absolute} protocol, both Alice and Bob
output $m$ qubits, and all the remaining qubits are
discarded. At the end of a \emph{conditional} protocol, one 
party sends a special symbol (either a $\suc$ or a
$\fail$) to the other party. If $\suc$ is sent, we say the protocol
\emph{succeeds}, and both party output $m$ qubits, as in the absolute
case. However, if $\fail$ is sent, we say the protocol
\emph{fails}. In this case, both party discard  
everything, and nothing is output. The special symbol isn't counted as
part of the communication complexity. If a protocol $\mathcal{P}$
outputs a state $\sigma$ over input $\rho$, we write this as
$\mathcal{P}(\rho)= \sigma$. If $\mathcal{P}$ is probabilistic, we
define its \emph{success probability} over $\rho$ to be the
probability that it succeeds if $\rho$ is the input, and denote this by
$P_{\mathcal{P}}[\rho]$. The \emph{ideal success probability} of a
protocol $\mathcal{P}$ is its success probability over the perfect EPR
pairs $\Psi_n$. A probabilistic EDP is \emph{ideal}, if its ideal
success probability is 1.
}

\section{The Measure-$r$ Model}
\label{sec:measure-r}

We prove an upper bound on the fidelity of 0-bit EDPs with respect to
the measure-$r$ error model.

\subsection{Notations and Definitions}

We start with more notations and definitions. 

A \emph{binary indicator vector}, often denoted by $\vc{v}$, 
 is an $n$-dimensional vector, whose each entry is an 
element from $\{0,1,*\}$. The \emph{degree} of a binary indicator
vector $\vc{v}$ is the number of entries that are not $*$,
and we write this as $\deg(\vc{v})$.  There are
$2^r\cdot{n\choose r}$ binary indicator vectors of degree $r$.  
Each binary indicator vector $\vc{v}$ corresponds to a unique
bipartite 
quantum state $\ket{\phi_{\vc{v}}}$ in ${\cal H}^{2^n}$ in the
following way: 

$$\ket{\phi_{\vc{v}}} = \bigotimes_{j=0}^{n-1}\ket{\phi_j},\;\;
\mbox{where}\;\;
\ket{\phi_j} = \left\{
\begin{array}{lll}
\ket{0}^A\ket{0}^B & & \mbox{if $v[j] = 0$}\\
\ket{1}^A\ket{1}^B & & \mbox{if $v[j] = 1$}\\
\Phi^+ & & \mbox{\sf if $v[j] = *$}\\
\end{array}
\right.
$$

The state $\ket{\phi_{\vc{v}}}$ is called an \emph{error state}, where
$\vc{v}$ is called its \emph{error indicator vector}. The
\emph{degree} of state $\ket{\phi_{\vc{v}}}$ is the degree of its
indicator vector. The error model for the measure-$r$ model, denoted
by $\mathcal{M}^{\mbox{\sf m}}_{n,r}$,  is defined
to be 
\begin{equation}
\mathcal{M}^{\mbox{\sf m}}_{n,r} =
\{\ket{\phi_{\vc{v}}}\;|\;\mbox{\sf  
$\vc{v}$ is an $n$-dimensional
 binary indicator such that $\deg(\vc{v}) = r$} \}
\end{equation}

An $n$-dimensional 0-1 vector $x$ is \emph{consistent} with a binary
indicator vector $\vc{v}$, if $x[j] = \vc{v}[j]$ for all $j$ such that
$\vc{v}[j] \ne *$. We write this as $x\cons \vc{v}$. For any $\vc{v}$
of degree $r$, there are $2^{n-r}$ 0-1 vectors $x$ consistent with
$\vc{v}$.  
It is not hard to verify that
\begin{equation}
\ket{\phi_{\vc{v}}} = {1\over2^{(n-r)/2}}\sum_{x\cons\vc{v}}
\ket{x}^A\ket{x}^B
\end{equation}

\subsection{Two Useful Lemmas}
We prove 2 lemmas that would be useful for the proofs in this
paper. Both lemmas are about how much ``deviation'' a quantum state
undergoes when applied various unitary operations. 

First, we consider the ``deviation'' of an arbitrary pure state under
the operations $\{I, X, Y, Z\}$ over its first qubit. We have the
following lemma:
\begin{lemma}
\label{lemma:unitary-singlebit}
Let $\ket{\phi}$ and $\ket{\psi}$ be two pure states of the same
dimension, not necessarily 
bipartite. Let $I$, $X$, $Y$, and $Z$ be the 
unitary operations over the first qubit of $\ket{\phi}$. Then we have
\begin{equation}
\sum_{U\in\{I,X,Y,Z\}}|\bra{\phi}U\ket{\psi}|^2\le 2
\end{equation}
\end{lemma}
\begin{proof}
\remove{
Consider the mixed state $\rho$ obtained by applying a random unitary
operation $U\in\{I,X,Y,Z\}$ over the first qubit of $\ket{\psi}$. In
other words, let 
$\rho  = {1\over4}\sum_{U\in\{I,X,Y,Z\}}U\dyad{\phi}{\phi}U^\dagger$.
Then we have
$$\sum_{U\in\{I,X,Y,Z\}}|\bra{\phi}U\ket{\phi}|^2 = 4F(\rho,
\dyad{\phi}{\phi}) $$
By the monotonicity of fidelity, the fidelity of $\rho$ and
$\dyad{\phi}{\phi}$ doesn't decrease if we trace out all but the first
qubits of $\rho$ and $\dyad{\phi}{\phi}$. However, tracing out all but
the first qubit will make $\rho$ the completely mixed state 
${I\over  2}$ and leave $\dyad{\phi}{\phi}$ a pure state. It is
clearly that the maximal fidelity of a pure state and the $I/2$ is
$1/2$. 
}
We write 
\begin{math}
\ket{\phi} = \alpha_0\ket{0}\ket{\phi_0}+\alpha_1\ket{1}\ket{\phi_1}
\end{math}
and
\begin{math}
\ket{\psi} = \beta_0\ket{0}\ket{\psi_0}+\beta_1\ket{1}\ket{\psi_1}
\end{math}

Then we have
\begin{eqnarray*}
\bra{\phi}I\ket{\psi} & = & \alpha_0^*\beta_0\braket{\phi_0}{\psi_0}
+\alpha_1^*\beta_1\braket{\phi_1}{\psi_1}\\
\bra{\phi}X\ket{\psi} & = & \alpha_1^*\beta_0\braket{\phi_1}{\psi_0}
+\alpha_0^*\beta_1\braket{\psi_0}{\phi_1} \\
\bra{\phi}Y\ket{\psi} & = &
-i\alpha_1\beta_0^*\braket{\phi_1}{\psi_0} + 
i\alpha_0\beta_1^*\braket{\phi_0}{\psi_1} \\
\bra{\phi}Z\ket{\psi} & = &
\alpha_0^*\beta_0{\phi_0}{\psi_0} - \alpha_1^*\beta_1
\braket{\phi_1}{\psi_1}
\end{eqnarray*}
Therefore
\begin{eqnarray*}
\sum_{U\in\{I,X,Y,Z\}}|\bra{\phi}U\ket{\psi}|^2 & = & 
2|\alpha_0\beta_0|^2|\braket{\phi_0}{\psi_0}|^2+
2|\alpha_1\beta_1|^2|\braket{\phi_1}{\psi_1}|^2 +
2|\alpha_0\beta_1|^2|\braket{\phi_0}{\psi_1}|^2 +
2|\alpha_1\beta_0|^2|\braket{\phi_1}{\psi_0}|^2\\
& \le & 2|\alpha_0|^2|\beta_0|^2+2|\alpha_1|^2|\beta_1|^2 +
2|\alpha_0|^2|\beta_1|^2 +2|\alpha_1|^2|\beta_0|^2\\
& = & 2(|\alpha_0|^2+|\alpha_1|^2)(|\beta_0|^2+|\beta_1|^2) \\
& = & 2
\end{eqnarray*}
\end{proof}

An immediate corollary is
\begin{corollary}
Let $\ket{\phi}$ be a pure sate. We have
$\sum_{U\in\{I,X,Y,Z\}}|\bra{\phi}U\ket{\phi}|^2\le 2$.
\end{corollary}

Next, we consider quantum states and operations over bipartite
systems. In particular, we study the ``deviation'' of a general
bipartite state under 
unitary operations of the form $U\otimes U^*$. We
interpret $U\otimes U^*$ as Alice applies $U$ to her first qubit and
Bob applies $U^*$ to his first qubit. Again, we consider 
$U\in\{I, X, Y, Z\}$.

We have the following lemma.
\begin{lemma}
\label{lemma:bell-fidelity}
Let $\ket{\phi}$ be a pure state in a bipartite system shared between
Alice and Bob.
Let $I$,  $X\otimes X^*$, $Y\otimes Y^*$, and 
$Z\otimes Z^*$ be the unitary operations over the first 
All these 4 operations work on the first qubit of Alice and the first
qubit of Bob.
Then we have
\begin{equation}
\braket{\phi}{\phi} + \bra{\phi}(X\otimes X^*)\ket{\phi} +
\bra{\phi}(Y\otimes Y^*)\ket{\phi} +
\bra{\phi}(Z\otimes Z^*)\ket{\phi} = 4\sfidel(\ket{\phi})
\end{equation}
\end{lemma}

\begin{proof}
We first consider how the Bell states behave under these unitary
operations. It is easy to verify the result, which we compile into the
following table.

\begin{center}
\begin{table}[h]
\label{table:bell}
\caption{The Bell States under operators}
\begin{center}
\begin{tabular}{|l|rrrr|}
\hline
\textbf{state} & $\Phi^+$ & $\Phi^-$ & $\Psi^+$ & $\Psi^-$ \\
\hline
$I\otimes I^*$ & $\Phi^+$ & $\Phi^-$ & $\Psi^+$ & $\Psi^-$ \\
$X\otimes X^*$ &  $\Phi^+$ & -$\Phi^-$ & $\Psi^+$ & -$\Psi^-$\\
$Y\otimes Y^*$ &  $\Phi^+$ & -$\Phi^-$ & -$\Psi^+$ & $\Psi^-$\\
$Z\otimes Z^*$ & $\Phi^+$ & $\Phi^-$ & -$\Psi^+$ & -$\Psi^-$\\
\hline
\end{tabular}
\end{center}
\end{table}
\end{center}
It is easy to see that the state $\Phi^+$ is invariant under any of
the 4 operations, while other Bell states will change their signs
under some operations.

Notice the 4 Bell states form an orthonormal basis for a bipartite
system of 2 qubits. 
We decompose
$\ket{\phi}$ into the Bell basis and write 
$$\ket{\phi} = \alpha_0\Phi^+\otimes \ket{\psi_0} +
\alpha_1\Phi^-\otimes \ket{\psi_1} +
\alpha_2\Psi^+\otimes \ket{\psi_2} +
\alpha_3\Psi^-\otimes \ket{\psi_3}
$$
where $\sum_{j=0}^3|\alpha_j|^2 = 1$. 
Therefore we have
\begin{eqnarray*}
\braket{\phi}{\phi} & = & 
|\alpha_0|^2 + |\alpha_1|^2 + |\alpha_2|^2 + |\alpha_3|^2\\
\bra{\phi}(X\otimes X^*)\ket{\phi}& = & 
|\alpha_0|^2 - |\alpha_1|^2 + |\alpha_2|^2 - |\alpha_3|^2\\
\bra{\phi}(Y\otimes Y^*)\ket{\phi}& = & 
|\alpha_0|^2 - |\alpha_1|^2 - |\alpha_2|^2 + |\alpha_3|^2\\
\bra{\phi}(Z\otimes Z^*)\ket{\phi}& = & 
|\alpha_0|^2 + |\alpha_1|^2 - |\alpha_2|^2 - |\alpha_3|^2\\
\end{eqnarray*}
and so, 
$$\braket{\phi}{\phi} + \bra{\phi}(X\otimes X^*)\ket{\phi} +
\bra{\phi}(Y\otimes Y^*)\ket{\phi} +
\bra{\phi}(Z\otimes Z^*)\ket{\phi} = 4|\alpha_0|^2 = 
4\sfidel(\ket{\phi})$$
\end{proof}

The above lemma implies an alternative definition of the base fidelity
of a pure state. 

\remove{

\subsection{A Positive Result}

We point out that the measure-$r$ error model is consistent with a
model commonly used in the quantum error correcting codes, namely, the
\emph{bit-flip} error model. In the bit-flip error model.

The following theorem is a corollary from~\cite{BDS+96}, and we
include it here for completeness. The proof can be found in
Appendix~\ref{app:positive}.
\begin{theorem}
\label{thm:pos-measure-r}
There exists a constant $c$ such that for any sufficiently large $n$ and
$r\le c\cdot n$, there exists a deterministic $t$-bit entanglement
distillation protocol 
of fidelity 1 under the measure-$r$ model.
\qed
\end{theorem}
}

\subsection{A Tight Bound for the 
No-Communication Case}

We prove that the fidelity of
0-bit EDPs for the measure-$r$ error model is at most $1-r/2n$, even
if the 
protocols are only required to output one qubit-pair. Notice that
fidelity is monotone. Therefore if no protocol can output a single
qubit pair of fidelity at least $1-r/2n$, then no protocol can output
multiple qubit pairs of fidelity at least $1-r/2n$.

\begin{theorem}
\label{thm:neg-measure-r}
For any probabilistic 0-bit protocol $\mathcal{P}$ that outputs
one qubit pair, we have $F(\mathcal{P}) \le 1-{r\over 2n}$ with
respect to the measure-$r$ model. 
\end{theorem}
Notice that there exists a very simple probabilistic 0-bit protocol
that has fidelity $1-{r\over 2n}$: Alice and Bob use their shared
random string to uniformly pick an EPR  pair and output it.
If this pair is measured, (which happens with
probability $r/n$), the fidelity is 1/2,  and
otherwise it is $1$. So the overall fidelity 
is exactly $1-r/2n$. So our upper bound is  tight.
\begin{proof}
We consider a slightly different error model, where a \emph{random}
$r$ out of $n$ EPR pairs are measured. This corresponds to the density
matrix 
$$\rho = {1\over2^n{n\choose r}}\sum_{\vc{v}: \deg\vc{v} = r}
\dyad{\phi_{\vc{v}}}{\phi_{\vc{v}}}$$ 
Notice that this is the ``average case'' version of the measure-$r$
model. Thus if we prove an upper bound on the fidelity of
$\mathcal{P}$ over $\rho$, then it is also an upper bound with respect
to the measure-$r$ model.

We shall prove that no \emph{deterministic} 0-bit protocol can have a
fidelity higher than $1-r/2n$ if $\rho$ is the input. Then, 
we conclude that no probabilistic protocol can
have a fidelity higher than $1-r/2n$, too, since fidelity is linear. 

Notice $\mathcal{P}$ doesn't involve any communication, we can model
it as Alice and Bob both applying a unitary operation to their share
of qubits, outputs the first qubit and discard the rest. 

Suppose the unitary operators of Alice and Bob are
$U_A$ and $U_B$. We denote the states under these operations by 
\begin{eqnarray*}
U_A\ket{x} & \longrightarrow & \ket{\phi_x} \\
U_B\ket{x} & \longrightarrow & \ket{\psi_x} 
\end{eqnarray*}
Notice that we use ``$\longrightarrow$'' instead of ``$=$'' since we
allow Alice and Bob to use ancillary bits. Clearly, the vectors
$\{\ket{\phi_x}\}_x$ are orthonormal, and so are the vectors
$\{\ket{\psi_x}\}_x$.

We shall prove that
\begin{equation}
\label{eqn:neg-measure-r-1}
{1\over2^r{n\choose r}}\sum_{\deg\vc{v} =r}
\left[[\sfidel((U_A\otimes U_B)
\dyad{\phi_{\vc{v}}}{\phi_{\vc{v}}}(U_A\otimes U_B)^\dagger)\right] 
\le 1-{r\over 2n},
\end{equation}
which shall imply our lemma.
By Lemma~\ref{lemma:bell-fidelity}, (\ref{eqn:neg-measure-r-1}) 
is equivalent to 
\begin{equation}
\label{eqn:sum}
{1\over2^r{n\choose r}}\sum_{\deg\vc{v} =r}
\left[\sum_{U\in\{I,X,Y,Z\}}
\bra{\phi_{\vc{v}}}(U_A\otimes U_B)^\dagger
(U\otimes U^*)(U_A\otimes U_B)\ket{\phi_{\vc{v}}}\right] \le 
4(1-{r\over2n})
\end{equation}

We expand the left hand side: 
Notice that 
$$(U_A\otimes U_B)\ket{\phi_{\vc{v}}}
= {1\over2^{(n-r)/2}}\sum_{x\cons\vc{v}}\ket{\phi_x}\ket{\psi_x}$$
and so we have
\begin{eqnarray*}
\bra{\phi_{\vc{v}}}(U_A\otimes U_B)^\dagger
(U\otimes U^*)(U_A\otimes U_B)\ket{\phi_{\vc{v}}}
& = & 
{1\over2^{n-r}}\sum_{x\cons\vc{v}}\sum_{y\cons\vc{v}}
\bra{\phi_x}U\ket{\phi_y}\cdot
\bra{\psi_x}U^*\ket{\psi_y}\\
\end{eqnarray*}
for any unitary operation $U$.
So, (\ref{eqn:sum}) is equivalent to 
\begin{equation}
\label{eqn:sum-2}
{1\over2^n{n\choose r}}\sum_{\deg\vc{v} =r}
\sum_{x\cons\vc{v}}\sum_{y\cons\vc{v}}\sum_{U\in\{I, X, Y, Z\}}
\bra{\phi_x}U\ket{\phi_y}\cdot
\bra{\psi_x}U^*\ket{\psi_y}
\le 4(1-{r\over2n})
\end{equation}

However, by Cauchy-Schwartz, we have
\begin{eqnarray*}
& & \sum_{\deg\vc{v} =r}
\sum_{x\cons\vc{v}}\sum_{y\cons\vc{v}}\sum_{U\in\{I, X, Y, Z\}}
\bra{\phi_x}U\ket{\phi_y}\cdot
\bra{\psi_x}U^*\ket{\psi_y}\\
& \le &
\left(\sum_{\deg\vc{v} =r}
\sum_{x\cons\vc{v}}\sum_{y\cons\vc{v}}\sum_{U\in\{I, X, Y, Z\}}
|\bra{\phi_x}U\ket{\phi_y}|^2
\right)^{1\over2}\cdot
\left(
\sum_{\deg\vc{v} =r}
\sum_{x\cons\vc{v}}\sum_{y\cons\vc{v}}\sum_{U\in\{I, X, Y, Z\}}
|\bra{\psi_x}U^*\ket{\psi_y}|^2
\right)^{1\over2}
\end{eqnarray*}
Next, we estimate the terms on the right hand side:
\begin{eqnarray*}
\sum_{\deg\vc{v} =r}
\sum_{x\cons\vc{v}}\sum_{y\cons\vc{v}}\sum_{U\in\{I, X, Y, Z\}}
|\bra{\phi_x}U\ket{\phi_y}|^2
& = & 
\sum_x\sum_y \sum_{U\in\{I, X, Y, Z\}}
|\bra{\phi_x}U\ket{\phi_y}|^2 \;
\sum_{\deg \vc{v}=r\::\;x_1\cons\vc{v}\land x_2\cons\vc{v}} 1
\end{eqnarray*}

Notice that since $\ket{\phi_x}$'s are all orthonormal, we have
$\sum_y|\bra{\phi_x}U\ket{\phi_y}|^2 \le 1$
for all $x$'s. Thus
$$\sum_x\sum_y\sum_{U\in\{I,X,Y,Z\}}|\bra{\phi_x}U\ket{\phi_x}|^2\le
2^{n+2} $$

For any $x$ and $y$, we have
$$\sum_{\deg \vc{v}=r\::\;x\cons\vc{v}\land y\cons\vc{v}} 1 =
{n-|x\oplus y|\choose n-r - |x\oplus y|}$$
The reason is simple: the only freedom for $\vc{v}$ is where to put
the $(n-r)$ $*$'s. But for every position $k$ such that $x[k]\ne
y[k]$, we have to have $\vc{v}[k]=*$. Then we still have
$(n-r-|x\oplus y|)$ $*$'s we can put anywhere. So if $x\ne y$, 
$$\sum_{\deg \vc{v}=r\::\;x\cons\vc{v}\land y\cons\vc{v}} 1
\le {n-1\choose n-r-1}$$
Also notice that by Lemma~\ref{lemma:unitary-singlebit}, we have
$\sum_{U\in\{I,X,Y,Z\}}|\bra{\phi_x}U\ket{\phi_x}|^2\le 2$ for any
$x$. 

Putting things together, we have
\begin{eqnarray*}
\sum_{\deg\vc{v} =r}
\sum_{x\cons\vc{v}}\sum_{y\cons\vc{v}}\sum_{U\in\{I, X, Y, Z\}}
|\bra{\phi_x}U\ket{\phi_y}|^2
& \le & 
{n\choose r}\cdot
\sum_{x}\sum_{U\in\{I,X,Y,Z\}}|\bra{\phi_x}U\ket{\phi_x}|^2 +  
{n-1\choose r-1}\cdot
\sum_{x\ne y}
\sum_{U\in\{I, X, Y, Z\}} 
|\bra{\phi_x}U\ket{\phi_y}|^2 \\
& = & \left[{n\choose r} - {n-1\choose r-1}\right]\cdot
\sum_{x}\sum_{U\in\{I,X,Y,Z\}}|\bra{\phi_x}U\ket{\phi_x}|^2 + \\
& & 
{n-1\choose r-1}\cdot
\sum_{x}\sum_y
\sum_{U\in\{I, X, Y, Z\}} 
|\bra{\phi_x}U\ket{\phi_y}|^2 \\
& = & 
\left[{n\choose r}-{n-1\choose r-1}\right]\cdot2^{n+1} +
{n-1\choose r-1}\cdot2^{n+2} \\
& = & 2^{n+2}{n\choose r}(1-{r\over 2n})
\end{eqnarray*}
Similarly, we have
$$
\sum_{\deg\vc{v} =r}
\sum_{x\cons\vc{v}}\sum_{y\cons\vc{v}}\sum_{U\in\{I, X, Y, Z\}}
|\bra{\psi_x}U^*\ket{\psi_y}|^2
\le 2^{n+2}{n\choose r}(1-{r\over 2n})
$$
too.

Thus we have
$$\sum_{\deg\vc{v} =r}
\sum_{x\cons\vc{v}}\sum_{y\cons\vc{v}}\sum_{U\in\{I, X, Y, Z\}}
\bra{\phi_x}U\ket{\phi_y}\cdot
\bra{\psi_x}U^*\ket{\psi_y}\le
2^{n+2}{n\choose r}(1-{r\over 2n})
$$
which proves (\ref{eqn:sum-2}).

\end{proof}

\section{The Depolarization Model}
\label{sec:depolar}

We prove an upper bound on the fidelity of 
0-bit EDPs with respect to the depolarization model.

\subsection{Notations and Definitions}

We give notations and definitions used in this section.

We first describe the depolarization channel. A depolarization channel
${\cal D}$ of 
parameter $p$ is a super-operator defined as~\cite{NC00}
$${\cal D}(\rho) = (1-p)\cdot\rho + p\cdot {I\over2}$$

In other words, this channel behaves in the following manner: with
probability $(1-p)$, it keeps the state untouched, and with
probability $p$, it replaces that with the completely mixed state.

It is not hard to verify that after passing the second qubit through
this channel, the state $\Phi^+$ becomes a mixed state 
$$\rho_p = (1-{3p\over4})\dyad{\Phi^+}{\Phi^+} + {p\over4}(
\dyad{\Phi^-}{\Phi^-} + \dyad{\Psi^+}{\Psi^+} +
\dyad{\Psi^-}{\Psi^-} )
$$

The depolarization error model of $n$ qubit pairs and parameter $n$,
denoted as
$\mathcal{M}^{\mbox{\sf d}}_{n,p}$, consists of a single state:
$\mathcal{M}^{\mbox{\sf d}}_{n,p} = \{\rho_p^{\otimes n}\}$.

\remove{
\subsection{A Positive Result}

The depolarization channel model was studied by Bennett
et. al.~\cite{BDS+96}, and one can easily derive the following result
from their paper. For completeness, we include a proof in 
Appendix~\ref{app:positive}.
\begin{theorem}
\label{thm:pos-depolarization}
There exists a constant $c$ such that for any sufficiently large $n$ and
$p\le c$, there exists a deterministic $t$-bit entanglement
distillation protocol 
of fidelity 1 under the depolarization model.
\qed
\end{theorem}
}

\subsection{An Almost-Tight Bound for the No-Communication
  Case}
We prove that the maximal fidelity of
0-bit EDPs for the depolarization error model is $1-p/2$, even if the
protocols are only required to output one qubit-pair.

\begin{theorem}
\label{thm:neg-depolarization}
For any probabilistic 0-bit protocol $\mathcal{P}$ that outputs
one qubit pair, we have $F(\mathcal{P}) \le 1-{p\over 2}$ with
respect to the depolarization model. 
\end{theorem}
There exists a very simple
deterministic 0-bit protocol that has fidelity $1-{3p\over4}$: Alice
and Bob simply output the first qubit pair. It is very easy to verify
that the fidelity of this protocol is
$1-{3p\over4}$. Therefore the bound in the theorem is almost-tight (by
a constant factor).

The proof to Theorem~\ref{thm:neg-depolarization} is very similar to
that to Theorem~\ref{thm:neg-measure-r}, except that it is more
technical. We 
postpone the proof to Appendix~\ref{app:fidelity}.

\section{The Fidelity Model}
\label{sec:fidelity}

We study the communication complexity of EDPs with respect to the
fidelity error model. 

First, we give the definition of the fidelity error model.
For a bipartite system of $n$ qubit pairs, we define the fidelity
error model of parameter $\epsilon$ to 
be the set of all bipartite systems of fidelity at least
$1-\epsilon$. We denote the error model by 
\begin{equation}
\mathcal{M}^{\mbox{\sf \small f}}_{n,\epsilon}
= \{\rho\:|\:\mbox{\sf $\rho$ has dimension $2^{2n}$ and $F(\rho)\ge
  1-\epsilon$}\}
\end{equation}

Notice that this error model is very different from the two previous
models we studied, since it provides much less information than the
previous one. As a comparison, notice that in the measure-$r$
model, all the error states have fidelity $1/2^r$, and
in the depolarization model, the fidelity of the input is
$(1-3p/4)^n$, 
both are very small. However, Alice and Bob have the
additional information about the \emph{structure} of the input states,
and are able to use the information to do very well.

\subsection{Two Useful Facts About Positive Operators}
We present two  useful facts about positive operators.
used in the rest of the paper.

For two positive operators $A$ and $B$, we say $A$ \emph{dominates}
$B$, if $A-B$ is still a positive operator, and we write this as 
$A\dom B$, or equivalently, $B\pv A$. 
 
\begin{claim}
\label{claim:monotone-pos-op}
For any positive super-operator $\mathcal{E}$ and any positive
operators $A$ and $B$,  if $A\dom B$,  then
$\mathcal{E}(A)\dom\mathcal{E}(B)$.
\qed
\end{claim}
This directly follows the fact that $\mathcal{E}$ is linear 
and preserves the positivity of operators: If $A-B$ is a positive
operator, then $\mathcal{E}(A)- \mathcal{E}(B)= \mathcal{E}(A-B)$ is
also a positive operator. 

\begin{claim}
\label{claim:dom-prob}
Let $\rho$ and $\sigma$ be density matrices such that 
$\rho\dom a\cdot \sigma$, for some positive number $a$.
For any
POVM $\{E_m\}$, let $p_m=\tr(\rho E_m)$ and and $q_m=\tr(\sigma E_m)$
be the probabilities the measurement result being $m$ for $\rho$ and
$\sigma$, respectively. Then we have $p_m\ge a\cdot q_m$.
\qed
\end{claim}
This is obvious, since we have
$p_m-a\cdot q_m =\tr((\rho - a\cdot \sigma)E_m)\ge 0$.

\subsection{Upper and Lower Bounds for the Fidelity Model}

Ambainis, Smith, and Yang~\cite{ASY02} 
proved that in the fidelity error model of parameter $\epsilon$ (which
they called the 
``general error model''), the maximal fidelity of a protocol is 
$1-\frac{2^m-2^k}{2^m}\frac{2^n}{2^n-1}\epsilon$.
if the protocol
has $n$ qubit pairs as input, $k$ perfect EPR pairs as auxiliary
input, and outputs $m$ qubit pairs. In a special case where $k=0$ (no
auxiliary input) and $m=1$ (only one pair is output), the maximal
fidelity is $1-{2^n\over2^n-1}{\epsilon\over2}<1-\epsilon/2$. In other
words, no ``interesting'' entanglement distillation protocols exist
for the fidelity error model. Their result is tight, in that they also
constructed a protocol, namely the ``Random Permutation Protocol'',
which achieves  
a fidelity of $1-\frac{2^m-2^k}{2^m}\frac{2^n}{2^n-1}\epsilon$. One
can slightly modify this protocol to completely eliminate
communication, and still maintain a high fidelity. In the original
construction of the random permutation protocol, communication is used
in 2 places. First, Alice and Bob communicate to agree on a common
random permutation. This part of communication is not needed for a
probabilistic protocol. Second, Alice and Bob communicate to check if
their measurements agree. We can modify the protocol by having Alice
and Bob always ``pretend'' that they measurements agree. A careful
analysis shows that this modification won't change the fidelity of the
protocol by much. In fact, we have the following theorem:
\begin{theorem}
\label{thm:pos-fidelity-no-comm}
There exists a probabilistic 0-bit entanglement distillation protocol
of fidelity $1-{2^n\over2^n-1}{\epsilon\over2}<1-\epsilon/2$ 
with respect to the fidelity model of parameter
$\epsilon$.
\qed
\end{theorem}



The situation for conditional fidelity is very different.
In fact, Ambainis et. al. proved that good
protocols exists with high \emph{conditional fidelity}. 
In particular,
the following result can be easily derived from~\cite{ASY02}:
\begin{theorem}[\cite{ASY02}]
\label{thm:pos-fidelity}
For every $n$ and $s<n$, there exists probabilistic 
$s$-bit entanglement distillation protocols of conditional fidelity
$1-2^{-s}/(1-\epsilon)$ with respect to the fidelity model of
parameter $\epsilon$.
\end{theorem}
\begin{sketch}
Consider the ``Simple Random Hash'' protocol in~\cite{ASY02}. 
The original construction for this protocol 
in~\cite{ASY02} has
$(2n+2)$ bits of 2-way communication. But a close examination 
reveals that 1 bit of 1-way communication suffices. In
the original construction, Alice sends $2n$ bits to Bob to establish
a common random string, which are not needed for a probabilistic
protocol. In the original protocol, Bob also sends 1 bit
of his measurement result back to Alice. This bit can also be
eliminated in our model, since we allow one player (normally Alice) to
output a special symbol at the end of the protocol.
We then repeat the simplified 1-bit protocol
for $s$ rounds sequentially, and 
obtain an $s$-bit protocol of conditional
fidelity $1-2^{-s}/(1-\epsilon)$.
\end{sketch}
Furthermore, the ``Simple Random Hash'' protocol only consists of
1-way communication. Also notice that this protocol is ideal, in that
if the input is the perfect EPR pairs $\Psi_n$, then the protocol
always succeeds.

Therefore, to achieve a conditional fidelity of $1-\delta$, only
$\log{1\over\delta} - \log(1-\epsilon)$ bits of communication
is needed in the fidelity error model. Next, we shall prove a lower
bound on the communication complexity.

\begin{theorem}
\label{thm:neg-fidelity}
For any probabilistic $s$-bit protocol of ideal success probability
$p$, its conditional fidelity is at 
most $1-\epsilon p/2^{s+1}$ with respect to the fidelity model of
parameter $\epsilon$.
\end{theorem}
Immediately from the theorem, we obtain a 
$\log ({1\over\delta})-\log({1\over\epsilon})-1$
lower bound on the communication complexity for ideal protocols of
conditional fidelity $1-\delta$. In the usual setting where $\epsilon$
is a constant, our lower bound matches the upper bound from
Theorem~\ref{thm:pos-fidelity}, up to an additive
constant. Interestingly, the theorem is proven for protocols that only
output 1 qubit pair. However, this lower bound is good enough in that
it matches the upper bound of the Simple Random Hash protocol, which
in fact outputs many qubit pairs. In this sense, the communication
complexity is ``oblivious'' of the yield of the EDPs. This is quite
counter-intuitive. 

\begin{proof}
WLOG we assume the protocol only outputs one qubit pair. 
Consider a particular input state 
\begin{equation}
\rho_0 = (1-\epsilon')\Psi_n 
+ \epsilon'\cdot {I\over2^{2n}}
\end{equation}
It is a mixture of the perfect EPR pairs $\Psi_n$ (with probability
$1-\epsilon'$) and the completely mixed state $I\over2^{2n}$ (with
probability $\epsilon'$).
Notice that 
$F({I\over2^{2n}}) = {1\over2^{2n}} $. 
So if we set $\epsilon' = {2^{2n}\over2^{2n}-1}\epsilon$, then we have
$F(\rho) = 1-\epsilon$.
We shall prove that no deterministic, $s$-bit protocol has fidelity
more than $1-2^{-(s+1)}\epsilon p$ over state $\rho_0$, which will
imply that no probabilistic protocol can have fidelity more than
$1-2^{-(s+1)}\epsilon p$, too.

We fix a deterministic protocol $\mathcal{P}$. WLOG, we assume it
proceeds in \emph{rounds}: in
each round, one of the two parties (Alice or Bob) applies a
super-operator $\mathcal{E}$ to his or her share of qubits, and then
sends one (classical) bit to the other party. The protocol consists of
$s$ rounds:  one bit is sent in each round. Finally, Alice outputs the
special symbol, determining if the protocol succeeds or fails.

To analyze the behavior of the protocol $\mathcal{P}$ over the input
$\rho_0$, we consider how $\mathcal{P}$ behaves over state $\Psi_n$
and state $I\over2^{2n}$, respectively. We use $p$ (resp. $q$) to
denote the probabilities that $\mathcal{P}$ succeeds over state
$\Psi_n$ (resp. $I\over2^{2n}$). Notice $p$ is in fact the
ideal success probability of protocol $\mathcal{P}$. 
Then it is easy to see that 
\begin{equation}
F^c(\mathcal{P}(\rho_0)) = {(1-\epsilon')p\cdot
  F^c(\mathcal{P}(\Psi_n)) +  
  \epsilon' q\cdot F^c(\mathcal{P}({I\over2^{2n}}))
\over (1-\epsilon')p + \epsilon' q}
\end{equation}

Notice that we always have 
$F^c(\mathcal{P}(\Psi_n))\le1$. Since ${I\over2^{2n}}$ is a
disentangled state, $\mathcal{P}({I\over2^{2n}})$ is also
disentangled. Therefore we have
$F^c(\mathcal{P}({I\over2^{2n}}))\le 1/2$ by
Lemma~\ref{lemma:disentangle-fidelity}. We shall prove that 

\begin{equation}
\label{eqn:neg-fidelity-main}
q\ge p^2/2^s,
\end{equation}
 which will imply that

\begin{equation}
F(\mathcal{P}(\rho_0)) \le {{(1-\epsilon')+\epsilon' p/2^{s+1}}\over
{(1-\epsilon')+\epsilon' p/2^s}} =
1-{\epsilon' p\over2^{s+1}(1-{{2^s\over2^s-1}}\epsilon' p)}
\le 1-\epsilon p/2^{s+1}
\end{equation}

\newcommand{\one}{\mbox{\rm \small I}}
\newcommand{\two}{\mbox{\rm \small II}}

Now we prove that $q\ge p^2/2^s$. We analyse 2 cases separately: in
case 
I, the state $\Psi_n$ is the input to the protocol; in case II, the
state $I\over2^{2n}$ is the input to the protocol.
For each case, we keep track of the local density matrices of Alice
and Bob. 
In case I, we use $\tau_k^{\one,A}$ and $\tau_k^{\one, B}$ to denote
the local density matrices of Alice and Bob after the $k$-th round; in
case II, we use $\tau_k^{\two,A}$ and $\tau_k^{\two, B}$,
respectively. For $k=0$, we define the $\tau_0^{\one,A}$,
$\tau_0^{\one,A}$, $\tau_0^{\two,A}$, and 
$\tau_0^{\two,A}$ to be the density matrices at the moment that
protocol starts.

We give more definitions: after the $k$-th round, there are $2^k$
possibilities depending on the first $k$ bits communicated.
For any binary string $t\in\bit^k$, we use $\sigma_t^{\one, A}$
(resp. $\sigma_t^{\one, B}$) to denote the local density matrix of 
Alice (resp. Bob) after the $k$-th round in case I, conditioned on
that  the first $k$ 
bits communicated so far are $t[0],t[1], ..., t[k-1]$. We use
$p_t^{\one}$ to denote the probability that
this happens (that the first $k$ bits are $t[0],t[1], ..., t[k-1]$).
Obviously we have 
$p_t^{\one} = p_{t\conc0}^{\one} + p_{t\conc1}^{\one}$ for any
$t\in\bit^k$. Furthermore, we have the following equalities
\begin{eqnarray}
\sum_{t\in\bit^k}p_t^{\one} & = & 1\\
\sum_{t\in\bit^k}p_t^{\one}\cdot
\sigma_t^{\one, A} & = &\tau_k^{\one, A}\\
\sum_{t\in\bit^k}p_t^{\one}\cdot
\sigma_t^{\one, B} & = &\tau_k^{\one, B}
\end{eqnarray}

We define $\sigma_t^{\two, A}$, $\sigma_t^{\two, B}$, 
and  $p_t^{\two}$ for case II, similarly.

We use $\emptystring$ to denote the empty string. So we have
$p_{\emptystring}^{\one} = p_{\emptystring}^{\two} = 1$.

One important observation is that when the protocol
starts, the local density matrices for Alice and Bob are identical in
both cases:
\begin{equation}
\label{eqn:init-identical}
\sigma_\emptystring^{\one, A} = \sigma_\emptystring^{\one, B} = 
\sigma_\emptystring^{\two, A} = \sigma_\emptystring^{\two, B} =
{I\over2^n} 
\end{equation}

When the protocol proceeds, the local density matrices in two cases
will become different, since the state $\Psi_n$ is an entangled state,
while $I\over2^{2n}$ is not. However, they cannot differ ``too far'',
as we shall prove in the following lemma:

\begin{lemma}
\label{lemma:divert}
For all $k=0,1,...,s-1$ and all $t\in\bit^k$, we have 
$p_t^{\one}\cdot \sigma_t^{\one, A}\pv \sigma_t^{\two, A}$
and
$p_t^{\one}\cdot \sigma_t^{\one, B}\pv \sigma_t^{\two, B}$.
\end{lemma}
\begin{proof}
By induction. The base case is obvious.
 Now the inductive case. 
Consider the situation at the end of the $k$-th round. Suppose the
first $k$ bits sent are $t[0], t[1], ..., t[k-1]$. WLOG we assume
that in the $(k+1)$-th round, Alice applies a super-operator
$\mathcal{E}$ to her share of qubits, and send one bit $a$ to
Bob. 

First we consider the density matrix for Alice. 
Notice that in general, $a$ is the result of the measurement from
$\mathcal{E}$. Therefore, we can ``split'' $\mathcal{E}$ into
two positive super-operators 
$\mathcal{E}_0$ and $\mathcal{E}_1$, such that
\begin{eqnarray}
\label{eqn:split-1}
\mathcal{E}_0(\sigma_t^{\one, A}) & =
 & {p_{t\conc  0}^{\one}\over p_t^{\one}}\cdot 
\sigma_{t\conc 0}^{\one, A}
 \\ 
\label{eqn:split-2}
\mathcal{E}_1(\sigma_t^{\one, A}) & =
 & {p_{t\conc  1}^{\one}\over p_t^{\one}}\cdot
\sigma_{t\conc 1}^{\one, A}
 \\ 
\label{eqn:split-3}
\mathcal{E}_0(\sigma_t^{\two, A}) & =
 & {p_{t\conc  0}^{\two}\over p_t^{\two}}\cdot
\sigma_{t\conc 0}^{\two, A}
 \\ 
\label{eqn:split-4}
\mathcal{E}_1(\sigma_t^{\two, A}) & =
 & {p_{t\conc  1}^{\two}\over p_t^{\two}}\cdot
\sigma_{t\conc 1}^{\two, A}
\end{eqnarray}
Intuitively, $\mathcal{E}_0$ corresponds to the case that $a=0$ is
sent, and $\mathcal{E}_1$ corresponds to the case that $a=1$ is sent.

By inductive hypothesis, we have
\begin{equation}
\label{eqn:IH-A}
p_t^{\one}\cdot \sigma_t^{\one, A} 
\pv \sigma_t^{\two, A}
\end{equation}
Combining (\ref{eqn:IH-A}), (\ref{eqn:split-1}) and
(\ref{eqn:split-3}) with Claim~\ref{claim:monotone-pos-op} yields 
that 
\begin{equation}
\label{eqn:inductive-0}
p_{t\conc 0}^{\one}\cdot \sigma_{t\conc 0}^{\one, A}
= \mathcal{E}_0(p_t^{\one}\cdot \sigma_t^{\one, A} )
\pv 
\mathcal{E}_0(\sigma_t^{\two, A})
=
 {p_{t\conc  0}^{\two}\over p_t^{\two}}\cdot\sigma_{t\conc 0}^{\two,
   A} \pv \sigma_{t\conc 0}^{\two,
   A}
\end{equation}
Combining (\ref{eqn:IH-A}), (\ref{eqn:split-2}) and
(\ref{eqn:split-4}) with Claim~\ref{claim:monotone-pos-op} yields 
that 
\begin{equation}
\label{eqn:inductive-1}
p_{t\conc 1}^{\one}\cdot \sigma_{t\conc 1}^{\one, A}
= \mathcal{E}_1(p_t^{\one}\cdot \sigma_t^{\one, A} )
\pv \mathcal{E}_1(\sigma_t^{\two, A})
=
{p_{t\conc  1}^{\two}\over p_t^{\two}}\cdot
\sigma_{t\conc 1}^{\two, A}
\pv\sigma_{t\conc 1}^{\two, A}
\end{equation}

Now we consider the local density matrix for Bob. In case I, the
qubits between Alice and Bob are entangled. Therefore, the bit Alice
sends to Bob carries some information about his state. In terms of
the density matrix, Bob's local density matrix will ``split'' from
$\sigma_t^{\one,B}$ to $\sigma_{t\conc 0}^{\one,B}$ and 
$\sigma_{t\conc 1}^{\one,B}$. Notice that Bob doesn't perform any
operation to his qubits, and thus we have
\begin{equation}
\label{eqn:split-bob-I}
\sigma_t^{\one,B} = {{p_{t\conc 0}^{\one}}\over
{p_{t}^{\one}}}\cdot \sigma_{t\conc
  0}^{\one,B} + 
{{p_{t\conc 1}^{\one}}\over{p_t^{\one}}}
\cdot \sigma_{t\conc
  1}^{\one,B} 
\end{equation}
In case II, the qubits between Alice and Bob are
disentangled. Therefore, the bit sent by Alice carries no information
about Bob's own state. Thus Bob's local density matrix remains 
unchanged. Thus we have

\begin{equation}
\label{eqn:split-bob-II}
\sigma_t^{\two,B} = \sigma_{t\conc
  0}^{\two,B} = \sigma_{t\conc 1}^{\two,B} 
\end{equation}

By inductive hypothesis, we have

\begin{equation}
\label{eqn:IH-B}
p_t^{\one}\cdot \sigma_t^{\one, B}\pv \sigma_t^{\two, B}
\end{equation}

Combining (\ref{eqn:split-bob-I}), (\ref{eqn:split-bob-II}), and
(\ref{eqn:IH-B}), we have
\begin{eqnarray}
p_{t\conc 0}^{\one}\cdot\sigma_{t\conc 0}^{\one, B} & \pv &
p_t^{\one}\cdot \sigma_{t}^{\one, B} \pv \sigma_t^{\two, B} =
 \sigma_{t\conc  0}^{\two,B}\\ 
p_{t\conc 1}^{\one}\cdot\sigma_{t\conc 1}^{\one, B} & \pv &
p_t^{\one}\cdot \sigma_{t}^{\one, B} \pv \sigma_t^{\two, B} =
 \sigma_{t\conc  1}^{\two,B}
\end{eqnarray}
So the inductive case is proved.
\end{proof}

Now we are ready to prove (\ref{eqn:neg-fidelity-main}).
After $s$ bits are send,
Alice will decide whether to succeed or fail. In case I, we use $r_t$
to denote the probability that Alice choose to succeed conditioned on
that the bits communicated are $t[0],t[1],..., t[s-1]$. Notice we have
$p_t^{\one}\cdot \sigma_t^{\one, A}\pv \sigma_t^{\two, A}$, and thus
by Lemma~\ref{lemma:divert}, we know that in case II, the success
probability is at least $p_t^{\one}\cdot r_t$.

Therefore, we have
\begin{eqnarray}
p & = & \sum_{t\in\bit^s}r_t\cdot p_{t}^{\one}\\
q & \ge & \sum_{t\in\bit^s}r_t\cdot p_{t}^{\one}\cdot p_{t}^{\one}
\end{eqnarray}

\remove{
Then we have  
$p = \sum_{t\in\bit^s}r_t\cdot p_{t}^{\one}$. Now since we have
\begin{equation}
p_t^{\one}\cdot \sigma_t^{\one, A}\pv \sigma_t^{\two, A}.
\end{equation}
On the other hand, by
Lemma~\ref{lemma:divert}, we know that the success probability of the
protocol in case II is at least $r_t\cdot p_{t}^{\one}$
in the case that the bit communicated are
$t[0],t[1],..., t[s-1]$.

$\sum_{t\in\bit^s}r_t\cdot \left(q_{t}^{\one}\right)^2$, by
Claim~\ref{claim:dom-prob}. 
}
which implies that
\begin{eqnarray}
q & \ge & \sum_{t\in\bit^s}r_t\cdot \left(p_{t}^{\one}\right)^2  \\
& \ge & {1\over2^s}\left(\sum_{t\in\bit^s}r_t\right)\cdot
\left[\sum_{t\in\bit^s}r_t\cdot \left(p_{t}^{\one}\right)^2\right]\\
& \ge & {1\over2^s}\left(\sum_{t\in\bit^s}r_t\cdot
  p_{t}^{\one}\right)^2 \\
& = & {p^2\over2^t}
\end{eqnarray}
This proves the theorem.
\end{proof}

\remove{
The next result implies that the bound in
Theorem~\ref{thm:neg-fidelity} is very tight for the no-communication
case.

\begin{theorem}
\label{thm:pos-no-comm}
There exists a probabilistic 0-bit entanglement distillation protocol
of fidelity $blah$ with respect to the fidelity model of parameter
$\epsilon$.
\end{theorem}
\begin{proof}

\end{proof}
}

\section{Conclusions and Future Work}
\label{sec:conclusion}

In this paper, we studied the classical communication complexity of
entanglement distillation protocols in the setting of 
\emph{incomplete information}, where the input states are mixed states
or prepared 
adversarially. We study on the \emph{precise} communication
complexity of the protocols, as opposed to the asymptotic
results. We also focus on the communication complexity of EDPs of the
\emph{minimal} requirement on yield, i.e., only 1 qubit pair is
required as output. To the best of our knowledge, this is the first
paper that studies classical communication complexity in the
incomplete information setting, and also the first one to study the
precise communication complexity. In our setting, many techniques
don't work any more, e.g., the Law of Large Numbers, the Central
Limit Theorem (both only works in the aggregated setting, where one
has many copies of the identical object), and the conversion from
multi-round protocols to a single-round protocol (it requires that
the input state is pure, and Alice and Bob have the complete
information about it).

We considered 3 error models of the input state, and proved 3
corresponding results. The first 2 results are the ``base cases'' for
the measure-$r$ and the depolarization models. The result upper-bound
the maximum possible fidelity of 0-bit EDPs (i.e., EDPs that don't
employ any communication). Interestingly, In this case, the trivial
protocols that outputs a random pair are already optimal (or
near-optimal, in the depolarization model). Despite of their simple
statement, these results seems non-trivial to prove. A technique in
the proof is an alternative definition of the fidelity of pure
states. The technique may have its independent interest. The third
result is an almost tight lower bound on the communication complexity
of EDPs with respect to the fidelity model. Interestingly, although
the lower bound is proven for protocols of minimal yield, it matches
the upper bound given by a specific protocol that has very high
yield. In this sense, the communication complexity seems to be
oblivious to the yield of EDPs. This observation is somewhat
surprising, since this is not the case for QECCs.

We view our paper as a first step toward the much greater project of
understanding the communication complexity of EDPs in general. We feel
that this paper opens much more open problems than the ones it
solved. We list some of the open problems that we feel interesting:

\begin{enumerate}
\item \textbf{More Lower Bounds}

Our first 2 results on the measure-$r$ models and the depolarization
model are indeed the ``base-case'' result, in that they only solved
the problem where there is no communication at all. What happens when
there is communication? In particular, in the measure-$r$ model, if
$r=1$ and $n\ge 3$, then there exists deterministic EDPs of fidelity
1. This contrasts with the results that the maximum fidelity of 0-bit
probabilistic EDPs is $1-1/2n$. What about 1-bit EDPs? 

\item \textbf{Tighter Lower Bounds}

Our result on EDPs with respect to the depolarization channel is not
tight: we managed to prove an $1-p/2$ upper bound on the fidelity of
0-bit EDPs, but the lower bound given by the trivial protocol is
$1-3p/4$. We conjecture that $1-3p/4$ is the right upper bound but was
unable to prove it.

\item \textbf{EDPs with Initial Entanglement}

Our paper didn't consider EDPs where Alice and Bob share some initial
entanglement (possible in the form of EPR pairs). How would the
initial entanglement affect the communication complexity?

\item \textbf{Deterministic vs. Probabilistic EDPs}

All the results in our paper are proven against probabilistic EDPs,
where Alice and Bob share a classical random tape. Can one prove
stronger results against deterministic EDPs? Is there a trade-off
between the amount of shared randomness used and the amount of
classical communication?

\end{enumerate}

\section*{Acknowledgment}
We wish to thank Adam Smith for useful discussions. We also thank
Michael Nielsen for pointing out several references that were
overlooked by us.

\appendix

\remove{
\section{Proofs to the Positive Results in the Measure-$r$ Model and
  the Depolarization Model}
\label{app:positive}

\begin{proof}
\textbf{[to Theorem~\ref{thm:pos-measure-r}]}

blah
\end{proof}
}

\section{Proofs to the Results for
  the Depolarization Model}
\label{app:fidelity}


\begin{proof}
\textbf{[to Theorem~\ref{thm:neg-depolarization}]}

Notice that by changing the basis, we can write the density
matrix, $\rho_p$, in another form:
$$\rho_p = (1-p)\cdot \dyad{\Phi^+}{\Phi^+} + {p\over4}\cdot(
\dyad{00}{00} + \dyad{01}{01} + \dyad{10}{10}+\dyad{11}{11}
)$$
which gives another interpretation of the depolarization model: 
each EPR pair, is kept intact with probability $(1-p)$, and is
replaced by a completely mixed state with probability $p$.

This observation leads us to consider a related error model, namely
the ``random-corrupt'' model. In a random-corrupt model of parameter
$r$, $r$ EPR pairs are 
randomly chosen from the $n$ pairs and are ``corrupted'' --- meaning
being replaced by 
the completely mixed state 
${1\over4}(\dyad{00}{00} + \dyad{01}{01} +
\dyad{10}{10}+\dyad{11}{11})$. 

It is easy to see that a depolarization error model of parameter $p$
is simply a mixture of the random-corrupt models, with probability 
${n\choose  r}p^r(1-p)^{n-r}$ being of parameter $r$.

We shall prove that the maximal fidelity of any 0-bit protocol over
the random-corrupt model of parameter $r$ is $1-r/2n$. This will
imply our theorem, since we have
$$\sum_{r=0}^n {n\choose  r}p^r(1-p)^{n-r}(1-{r\over 2n}) =
1-{p\over2}.$$

\remove{
We first consider a 
We shall prove a stronger statement,
that is, for any 0-bit protocol, its average fidelity is at most
$1-3r/4n$, if the input is the $r$ EPR pairs corrupted. This is
stronger since this result will immediately imply the theorem:
$$\sum_{r=0}^n {n\choose  r}p^r(1-p)^{n-r}(1-{3r\over 4n}) =
1-{3p\over4}$$
}

As in the proof to Theorem~\ref{thm:neg-measure-r}, we only consider
deterministic protocols. 

We present more notations and definitions.
As \emph{extended indicator vector}, often denoted by $\vc{u}$,
 is an $n$-dimensional vector,
whose each entry is an element from $\{00,01,10,11, *\}$. Its
\emph{degree} is the number of entries that are not $*$. There are
$4^r{n\choose r}$ extended indicator vectors of degree $r$. Each
extended indicator vector $\vc{u}$ corresponds to a unique bipartite
state $\ket{\psi_{\vc{u}}}$ in the following way:

$$\ket{\psi_{\vc{u}}} = \bigotimes_{j=0}^{n-1}\ket{\phi_j},\;\;
\mbox{where}\;\;
\ket{\phi_j} = \left\{
\begin{array}{lll}
\ket{0}^A\ket{0}^B & & \mbox{if $v[j] = 00$}\\
\ket{0}^A\ket{1}^B & & \mbox{if $v[j] = 11$}\\
\ket{1}^A\ket{0}^B & & \mbox{if $v[j] = 10$}\\
\ket{1}^A\ket{1}^B & & \mbox{if $v[j] = 11$}\\
\Phi^+ & & \mbox{\sf if $v[j] = *$}\\
\end{array}
\right.
$$
We call such an $\ket{\psi_{\vc{u}}}$ an \emph{extended error state}.

An $2n$-dimensional 0-1 vector $x$  is \emph{consistent} with an
extended indicator vector $\vc{u}$, if $x[j]\conc x[n+j] = \vc{u}[j]$
for all $j$ such that $\vc{v}[j] \ne *$, and $x[j]=x[n+j]$ for all $j$
such that $\vc{v}[j] = *$.  
We write this as $x\cons \vc{u}$. There are $2^{n-r}$ 0-1 vectors $x$
consistent with an indicator vector of degree $r$. 
We view $x$ as the concatenation of
2 $n$-dimension vectors: $x = l\conc r$, and we write them as  
$l=\ll(x)$ and $r=\rr(x)$. 

With the notations, we can write the extended error states as
\begin{equation}
\psi_{\vc{u}} = {1\over
  2^{(n-r)/2}}\sum_{x\cons\vc{u}}\ket{\ll(x)}^A\ket{\rr(x)}^B 
\end{equation}

We define the \emph{discrepancy} of $x$ to
be $\disc(x)= \ll(s)\oplus \rr(s)$, where $\oplus$ stands for bit-wise
XOR. 
The \emph{degree of  discrepancy} of $x$ is $|\disc(x)|$, the Hamming
weight of $\disc(x)$. 
Clearly, there are ${n\choose d}2^{n}$ 0-1 vectors of dimension $2n$
having degree of discrepancy $d$.
Furthermore, if $x$ has degree of discrepancy $d$, then the number of
degree-$r$ extended indicator vectors $\vc{u}$ such that
$x\cons\vc{u}$ is 
${n-d\choose r-d}$. This is because for every $j$ such that 
$x[j]\ne x[n+j]$, we must have $\vc{u}[j] = x[j]\conc x[n+j]$ in order
to have $x\cons \vc{u}$. So the only freedom for $\vc{u}$ is to put
$(n-r)$ $*$'s in the $n-d$ places where $x[j] = x[n+j]$. 

Now we consider an arbitrary 0-bit protocol. We model it as Alice and
Bob both applying a unitary operation to their share 
of qubits, outputs the first qubit and discard the rest. 
Suppose the unitary operators of Alice and Bob are
$U_A$ and $U_B$. We denote the states under these operations by 
\begin{eqnarray*}
U_A\ket{x} & \longrightarrow & \ket{\phi_x} \\
U_B\ket{x} & \longrightarrow & \ket{\psi_x} 
\end{eqnarray*}

Then as in the proof to Theorem~\ref{thm:neg-measure-r}, we shall
prove that  

\begin{equation}
\label{eqn:sum-general}
{1\over4^r{n\choose r}}\sum_{\deg\vc{u} =r}
\left[\sum_{U\in\{I,X,Y,Z\}}
\bra{\psi_{\vc{u}}}(U_A\otimes U_B)^\dagger
(U\otimes U^*)(U_A\otimes U_B)\ket{\psi_{\vc{u}}}\right] \le 
4(1-{r\over2n})
\end{equation}
which will imply our theorem.

Notice that 
$$(U_A\otimes U_B)\ket{\psi_{\vc{u}}}
= {1\over2^{(n-r)/2}}\sum_{x\cons\vc{u}}\ket{\phi_{\ll(x)}}
\ket{\psi_{\rr(x)}}$$
and so we have
\begin{eqnarray*}
\bra{\psi_{\vc{u}}}(U_A\otimes U_B)^\dagger
(U\otimes U^*)(U_A\otimes U_B)\ket{\psi_{\vc{u}}}
& = & 
{1\over2^{n-r}}\sum_{x\cons\vc{u}}\sum_{y\cons\vc{u}}
\bra{\phi_{\ll(x)}}U\ket{\phi_{\ll(y)}}\cdot
\bra{\psi_{\rr(x)}}U^*\ket{\psi_{\rr(y)}}\\
\end{eqnarray*}

So we only need to prove that
\begin{equation}
\label{eqn:sum-general-2}
{1\over2^{n+r}{n\choose r}}\sum_{\deg\vc{u} =r}
\sum_{x\cons\vc{u}}\sum_{y\cons\vc{u}}\sum_{U\in\{I, X, Y, Z\}}
\bra{\phi_{\ll(x)}}U\ket{\phi_{\ll(y)}}\cdot
\bra{\psi_{\rr(x)}}U^*\ket{\psi_{\rr(y)}}
\le 4(1-{r\over2n})
\end{equation}

By Cauchy-Schwartz, we have
\begin{eqnarray*}
& & \sum_{\deg\vc{u} =r}
\sum_{x\cons\vc{u}}\sum_{y\cons\vc{u}}\sum_{U\in\{I, X, Y, Z\}}
\bra{\phi_{\ll(x)}}U\ket{\phi_{\ll(y)}}\cdot
\bra{\psi_{\rr(x)}}U^*\ket{\psi_{\rr(y)}}\\
& \le &
\left(\sum_{\deg\vc{u} =r}
\sum_{x\cons\vc{u}}\sum_{y\cons\vc{u}}\sum_{U\in\{I, X, Y, Z\}}
|\bra{\phi_{\ll(x)}}U\ket{\phi_{\ll(y)}}|^2
\right)^{1\over2}\cdot \\
& & 
\left(
\sum_{\deg\vc{u} =r}
\sum_{x\cons\vc{u}}\sum_{y\cons\vc{u}}\sum_{U\in\{I, X, Y, Z\}}
|\bra{\psi_{\rr(x)}}U^*\ket{\psi_{\rr(y)}}|^2
\right)^{1\over2}
\end{eqnarray*}

\remove{
===========================================

We have a problem here: we can only prove a $1-r/2n$ bound, instead of
$1-3r/4n$.

===========================================
}

Now we estimate 
\begin{eqnarray*}
\sum_{\deg\vc{u} =r}
\sum_{x\cons\vc{u}}\sum_{y\cons\vc{u}}\sum_{U\in\{I, X, Y, Z\}}
|\bra{\phi_{\ll(x)}}U\ket{\phi_{\ll(y)}}|^2
\end{eqnarray*}

Notice we can write $x$ as $x= \ll(x)\conc (\ll(x)\oplus\disc(x))$ and
$y$ as  $y= \ll(y)\conc (\ll(y)\oplus\disc(y))$. If there exists an
extended indicator vector $\vc{u}$ such that $x\cons\vc{u}$ and
$y\cons\vc{u}$, we must have $\disc(x) = \disc(y)$. This is because
that for every $j$ 
such that $\disc(x)[j]=1$, $x[j]$ and $x[n+j]$ differ. Thus we must
have $\vc{v}[j] = x[j]\conc x[n+j]$, which implies that 
$\vc{v}[j] = y[j]\conc y[n+j]$, and $\disc(y)[j]=1$. In fact, for
every $j$ such that $\disc(x)[j] = 1$, we have $x[j] = y[j]$ and
$x[n+j] = y[n+j]$. 

So we have

\begin{eqnarray*}
& & 
\sum_{\deg\vc{u} =r}
\sum_{x\cons\vc{u}}\sum_{y\cons\vc{u}}\sum_{U\in\{I, X, Y, Z\}}
|\bra{\phi_{\ll(x)}}U\ket{\phi_{\ll(y)}}|^2 \\
& = & 
\sum_{a\in\bit^n}\sum_{b\in\bit^n}
\sum_{U\in\{I, X, Y, Z\}}
|\bra{\phi_{a}}U\ket{\phi_{b}}|^2
\sum_{c\in\bit^n} \sum_{\deg \vc{u}=r:\;
[(a\conc(a\oplus c))\cons\vc{u}]\land 
[(b\conc(b\oplus c))\cons\vc{u}]}1 
\end{eqnarray*}
by a substituting $a$ for $\ll(x)$, $b$ for $\ll(y)$, and $c$ for
$\disc(x)$. 

Now we fix $a$ and $b$, and compute
$$\sum_{c\in\bit^n} \sum_{\deg \vc{u}=r:\;
[(a\conc(a\oplus c))\cons\vc{u}]\land 
[(b\conc(b\oplus c))\cons\vc{u}] }1
$$
We define $k=|a\oplus b|$.
For every $j$ where
$a[j]\ne b[j]$, we must have $c[j]=0$ and $\vc{u}[j]= *$. For every
$j$ where $a[j]=b[j]$, if we have $c[j] = 1$, then we must 
$\vc{u}[j]=a[j]\conc(a[j]\oplus 1)$; if we have $c[j] = 0$, then
$\vc{u}$ can be either $a[j]\conc a[j]$ or $*$. Therefore, 
of $n-k$ positions where $a[j]=b[j]$, $r$ would be chosen where
$vc{u}$ has a non-$*$ entry. Of these $r$ places, one has the freedom
to choose $c[j]=0$ or $c[j]=1$. For all other places, $c[j]=0$ and
$\vc{u}=*$. So we have
$$\sum_{c\in\bit^n} \sum_{\deg \vc{u}=r:\;
[(a\conc(a\oplus c))\cons\vc{u}]\land 
[(b\conc(b\oplus c))\cons\vc{u}]} 1
 = 2^r\cdot{n-k\choose r}$$
In other words,
\begin{equation}
\sum_{\deg\vc{u} =r}
\sum_{x\cons\vc{u}}\sum_{y\cons\vc{u}}\sum_{U\in\{I, X, Y, Z\}}
|\bra{\phi_{\ll(x)}}U\ket{\phi_{\ll(y)}}|^2 
= \sum_{a\in\bit^n}\sum_{b\in\bit^n}
\sum_{U\in\{I, X, Y, Z\}}
|\bra{\phi_{a}}U\ket{\phi_{b}}|^2 \cdot 2^r\cdot{n-|a\oplus b|\choose
  r} 
\end{equation}

Since $\ket{\phi_a}$'s are orthogonal, we have
$$\sum_{a}\sum_{b}\sum_{U\in\{I,X,Y,Z\}}
|\bra{\phi_{a}}U\ket{\phi_{b}}|^2 \le 2^{n+2}$$

Also by Lemma~\ref{lemma:unitary-singlebit}, we have
$$\sum_{a}|\bra{\phi_{a}}U\ket{\phi_{a}}|^2 \le 2^{n+1}$$

Therefore
\begin{eqnarray*}
& & 
\sum_{a\in\bit^n}\sum_{b\in\bit^n}
\sum_{U\in\{I, X, Y, Z\}}
|\bra{\phi_{a}}U\ket{\phi_{b}}|^2 \cdot 2^r\cdot{n-|a\oplus b|\choose
  r} \\
& \le & 
\sum_{a}|\bra{\phi_{a}}U\ket{\phi_{a}}|^2 \cdot 2^r
\left[{n\choose r}- {n-1\choose r}\right] +
2^r{n-1\choose r}\sum_{a\in\bit^n}\sum_{b\in\bit^n}
\sum_{U\in\{I, X, Y, Z\}}
|\bra{\phi_{a}}U\ket{\phi_{b}}|^2 \\
& \le & 2^{n+r+1}\left[{n\choose r}- {n-1\choose r}\right] +
2^{n+r+2}{n-1\choose r} \\
& = & 2^{n+r+2}{n\choose r}(1-{r\over 2n})
\end{eqnarray*}

which implies (\ref{eqn:sum-general-2}), which implies the theorem.
\end{proof}

\end{document}